\documentclass{PoS}
\usepackage{wrapfig}
\usepackage{rotating}
\usepackage{multirow}

\title{{\small{DESY 16-121, DO-TH 16/25}}
\\
The new ABMP16 PDFs}

\ShortTitle{The new ABMP16 PDFs}

\author{\speaker{Sergey Alekhin}%
         \thanks{This work was supported in part by the
European Commission through contract PITN-GA-2012-316704 ({HIGGS\-TOOLS}).
}\\ 
II. Institut f\"ur Theoretische Physik, Universit\"at Hamburg,
    Luruper Chaussee 149, D-22761 Hamburg, Germany;\\
        Institute for High Energy Physics,142281 Protvino, Russia\\
        E-mail: \email{sergey.alekhin@desy.de}}

\author{Johannes Bl\"umlein\\ 
        Deutsches Elektronensynchrotron DESY, Platanenallee 6, D--15738 Zeuthen, Germany\\
        E-mail: \email{Johannes.Bluemlein@desy.de}}

\author{Sven-Olaf Moch\\
II. Institut f\"ur Theoretische Physik, Universit\"at Hamburg,
    Luruper Chaussee 149, D-22761 Hamburg, Germany \\      
 E-mail: \email{sven-olaf.moch@desy.de}}

\author{Ringaile Pla\v cakyt\. e \\
   Deutsches Elektronensynchrotron DESY, 
   Notkestra{\ss}e 85, D--22607 Hamburg, Germany\\
        E-mail: \email{ringaile.placakyte@desy.de}}

\abstract{We present an update of the ABM12 PDF analysis including improved 
constraints due to the final version of the inclusive DIS HERA data, 
the Tevatron and LHC data on the $W$- and $Z$-production and those on heavy-quark production in 
the electron- and neutrino-induced DIS at HERA and the fixed-target experiments 
NOMAD and CHORUS. We also check the impact of the Tevatron and LHC 
top-quark production data on the PDFs and the strong coupling constant. We obtain 
$\alpha_s(M_Z)=0.1145(9)$ and 0.1147(8) with and without the 
top-quark data included, respectively. 
}

\FullConference{XXIV International Workshop on Deep-Inelastic Scattering and Related Subjects\\
		11-15 April, 2016\\
		DESY Hamburg, Germany}

\begin{document}

\begin{figure}
  \centering
  \includegraphics[width=0.48\textwidth,height=0.42\textwidth]{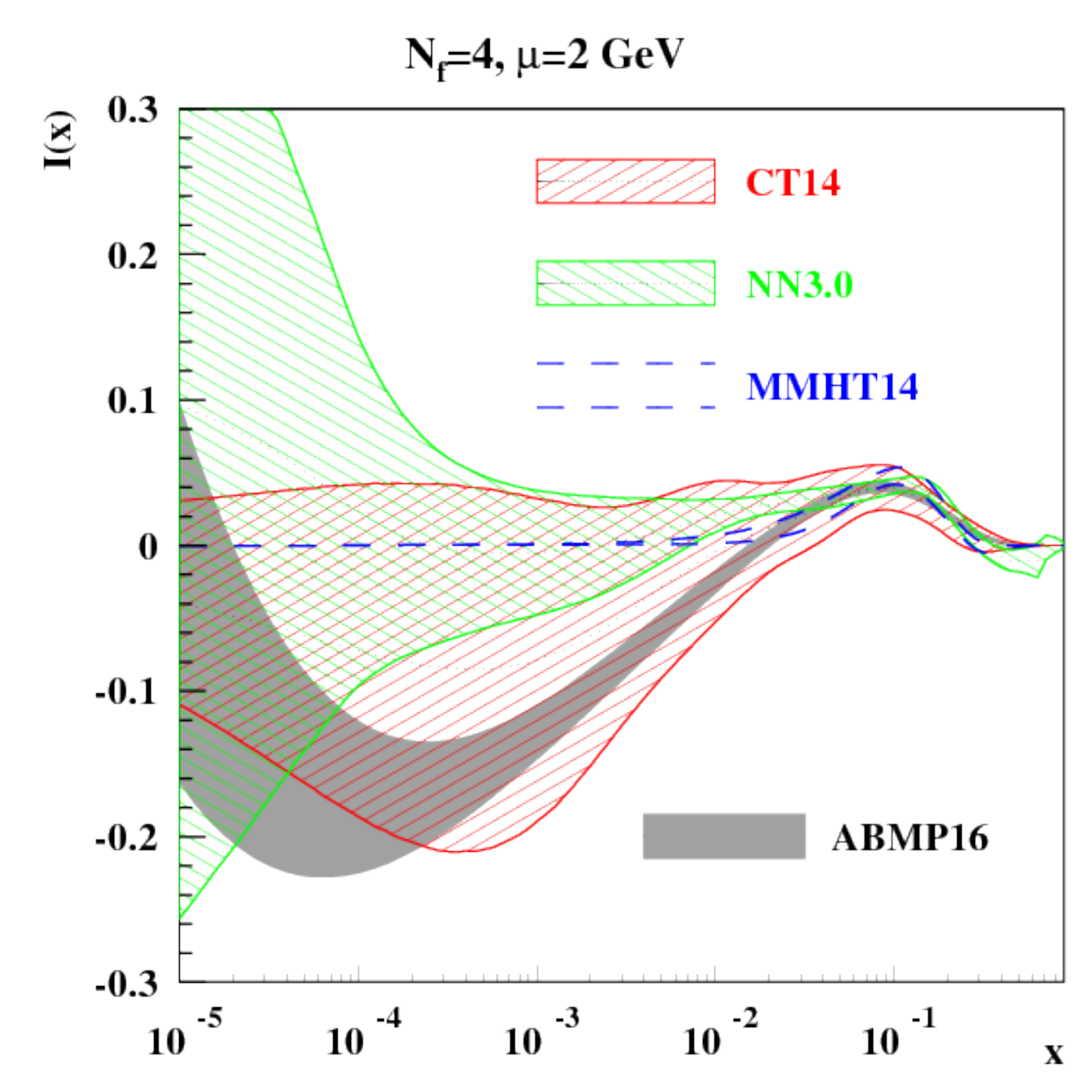}
  \includegraphics[width=0.47\textwidth,height=0.42\textwidth]{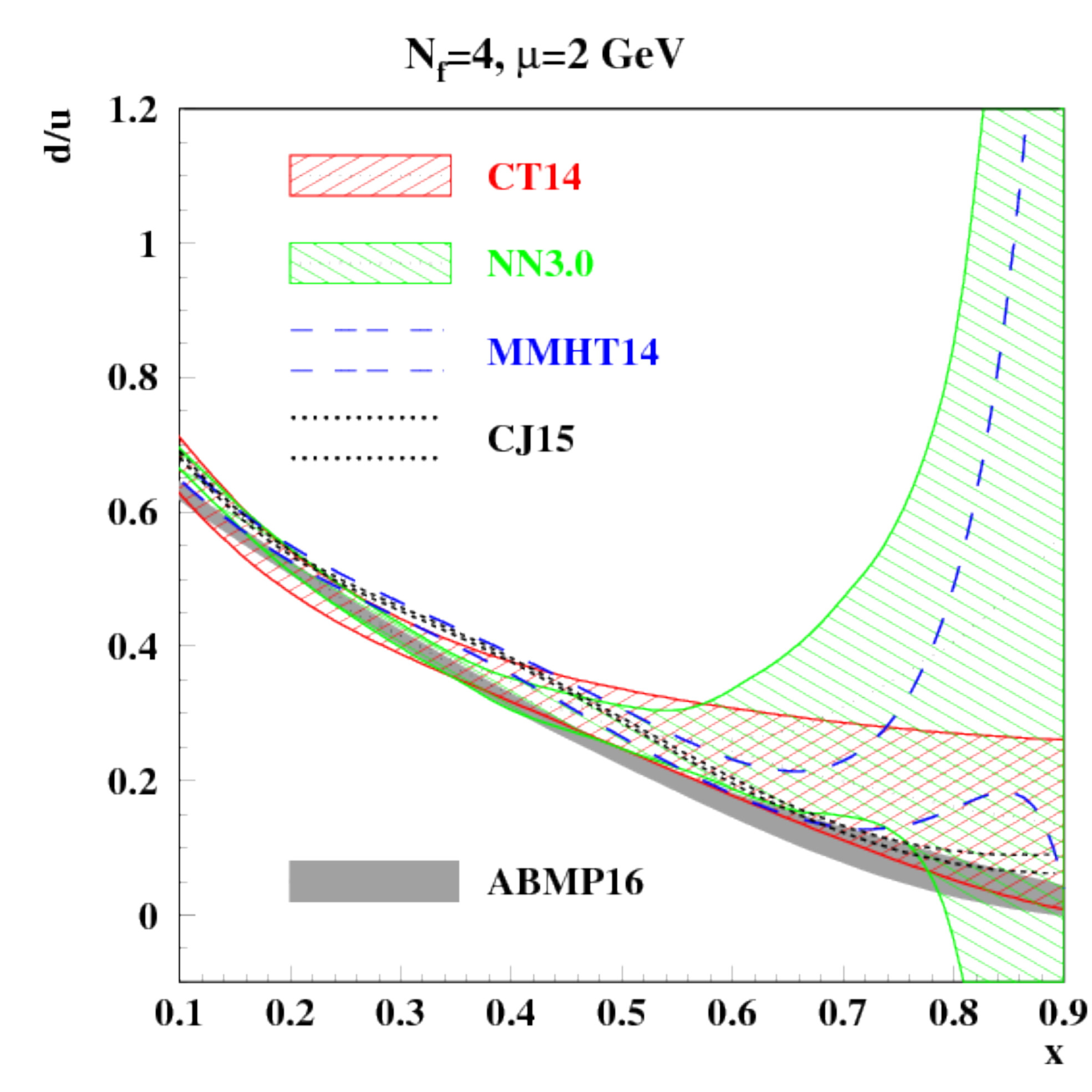}
  \caption{\small
Left: The $1\sigma$ error band for 
  the NNLO iso-spin asymmetry of the sea $I(x)$ for the 4-flavor scheme
    at the scale of
    $\mu=2~{\rm GeV}$ as a function of the Bjorken $x$ obtained in the 
present fit (gray shaded area) 
in comparison with the corresponding ones obtained in the 
CT14~\cite{Dulat:2015mca} (red right-tilted hatch), 
MMHT14~\cite{Harland-Lang:2014zoa} (blue dashed lines), and
NNPDF3.0~\cite{Ball:2014uwa} (green left-tilted hatch) analyses.
Right: The same for the ratio $d/u$ with the NLO CJ15 
results~\cite{Accardi:2016qay} added. 
}
    \label{fig:pdfs}
\end{figure}

A steady progress in accumulating and processing hard-scattering 
data from Tevatron, HERA, and the LHC stimulates an improvement in determining 
parton distribution functions (PDFs). In particular, a bulk 
of the data on $W$- and $Z$-boson production obtained recently 
at the Tevatron collider and the LHC opens a realistic perspective 
of using the Drell-Yan (DY) data as a powerful constraint to disentangle quark flavors.
The overall DY statistics accumulated by the Tevatron and LHC experiments
provides a typical uncertainty in the data of O(1\%) that is competitive 
to the most accurate deep-inelastic-scattering (DIS) data sets available.  
The DY data at large $W$- and $Z$-boson rapidity are particularly 
suitable to study the PDFs at small and large values of Bjorken variable $x$ 
probed in case of asymmetric kinematics. 
These benefits are employed in the updated version of the ABM12 PDF 
analysis~\cite{Alekhin:2013nda}
 including the most accurate DY data from Tevatron 
and the LHC. These data are reasonably well accommodated 
in the ABM12 fit, with a typical 
value of $\chi^2\lesssim1.5$ per data point, cf. Table~\ref{tab:dydata}. 
In combination with the existing 
DIS data from HERA and the fixed-target experiments they allow to 
separate the contribution from $u$- and $d$-quarks in a wide range of 
$x\simeq 10^{-4}\div 0.9$. 
This input is particularly useful for the extraction of the  
$d$-quark distribution at large $x$ 
since it allows to avoid uncertainties due to the 
modeling of nuclear effects, which appear in case of employing 
the DIS deuteron data for this purpose. The statistical
accuracy of the $d$-quark distribution, which can be achieved using the DY data 
is comparable to the one for the existing DIS deuteron data sets 
used earlier in the ABM12 fit~\cite{Alekhin:2015cza}. Therefore, 
in the present analysis we skip the latter in order to reduce 
theoretical uncertainties keeping the overall one on the same 
level, cf. Fig.~\ref{fig:pdfs}.
The forward DY data also help to disentangle the sea quark distributions 
at small $x$. In particular this concerns the shape of the isospin asymmetry 
$I(x)=x[\bar d(x) - \bar u(x)]$ now parameterized in a 
model-independent way allowing to release a constraint 
$I(x) \sim x^{0.7}$ imposed in the ABM12 fit. 
This constraint has been motivated by Regge-phenomenology arguments valid 
asymptotically for $x\rightarrow 0$. 
However, an explicit onset of this asymptotic is not easily specified and has
to be extracted from the experimental 
data. Releasing the Regge-like constraint on $I(x)$, 
we observe a substantial deviation 
from 0 at $x\sim 10^{-4}$ and a turnover of this trend at smaller
$x$ that still allows for a Regge-like shape at $x\sim 10^{-6}$. 
Determinations of the sea iso-spin asymmetry obtained in
the CT14~\cite{Dulat:2015mca}, MMHT14~\cite{Harland-Lang:2014zoa},
and NNPDF3.0~\cite{Ball:2014uwa} PDF fits
are in a broad agreement with our results although the uncertainties  
obtained by other groups are much bigger than ours 
because of fewer data used, cf. Fig.~\ref{fig:pdfs} and 
Table~\ref{tab:dydata}. The strange quark distribution 
determined from a global PDF fit 
is commonly separated from the non-strange 
ones using the data on charm production in neutrino-induced DIS. 
In the present analysis the strange sea is refined as compared to ABM12 
due to the addition of the recent NOMAD and CHORUS data. The former 
allows to pin down the strange sea at large $x$, where other 
experiments are not conclusive and the latter provides a unique 
information about the charm-production dynamics, 
being not sensitive to details of modeling its fragmentation~\cite{Alekhin:2014sya}. 
The resulting strange sea suppression factor updated is 
$\kappa(20~{\rm GeV}^2)=0.658\pm0.026$ for the 3-flavor PDFs. Furthermore 
the updated PDF predictions were compared to the ATLAS and CMS data on
associated '$W$+charm' production and good agreement was found 
at NLO in QCD. 

\begin{figure}
\begin{minipage}[c]{0.62\textwidth}
\includegraphics[width=0.85\textwidth, angle=0]{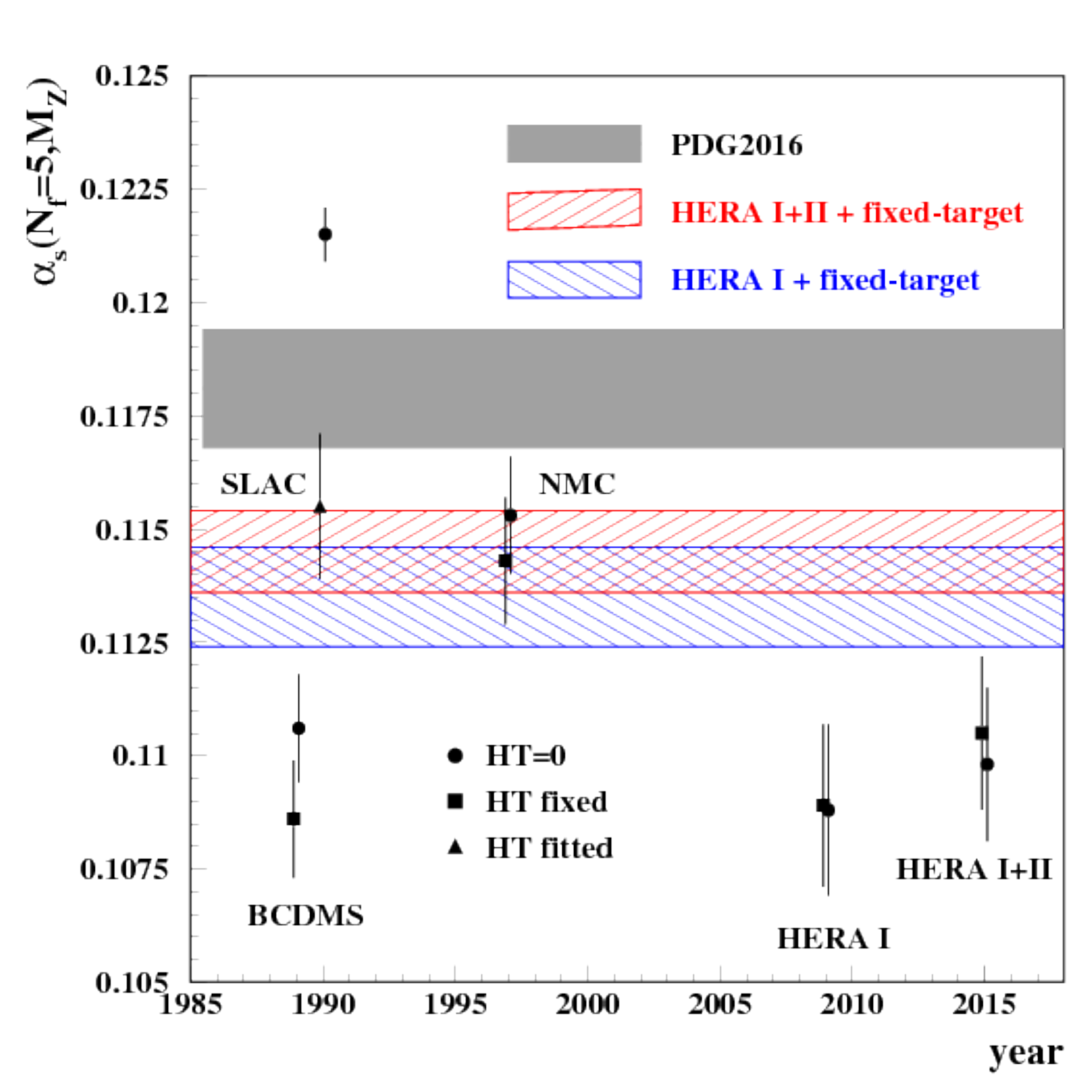}
  \end{minipage}\hfill
  \begin{minipage}[c]{0.36\textwidth}
\caption{\small
The value of $\alpha_s$ preferred by various DIS data samples 
employed in present analysis w.r.t. the year of data publication. Three variants
of the fit with different treatment of the HT terms are presented: 
HT set to 0 or to the ones obtained in the combined fit 
(circles and squares, respectively) and fitted to the one particular data set 
(triangles). 
The $\alpha_s$ bands obtained by using the combination of the fixed-target SLAC, BCDMS,
and NMC samples with the ones from the HERA Run-I (left-tilted hatches) and 
the Run-I+II (right-tilted hatches) as well as the PDG 
average~\cite{Agashe:2014kda} are given for 
comparison.  }
\label{fig:alps}
\end{minipage}
\end{figure}
\begin{figure}
  \centering
  \includegraphics[width=0.97\textwidth,height=0.37\textwidth]{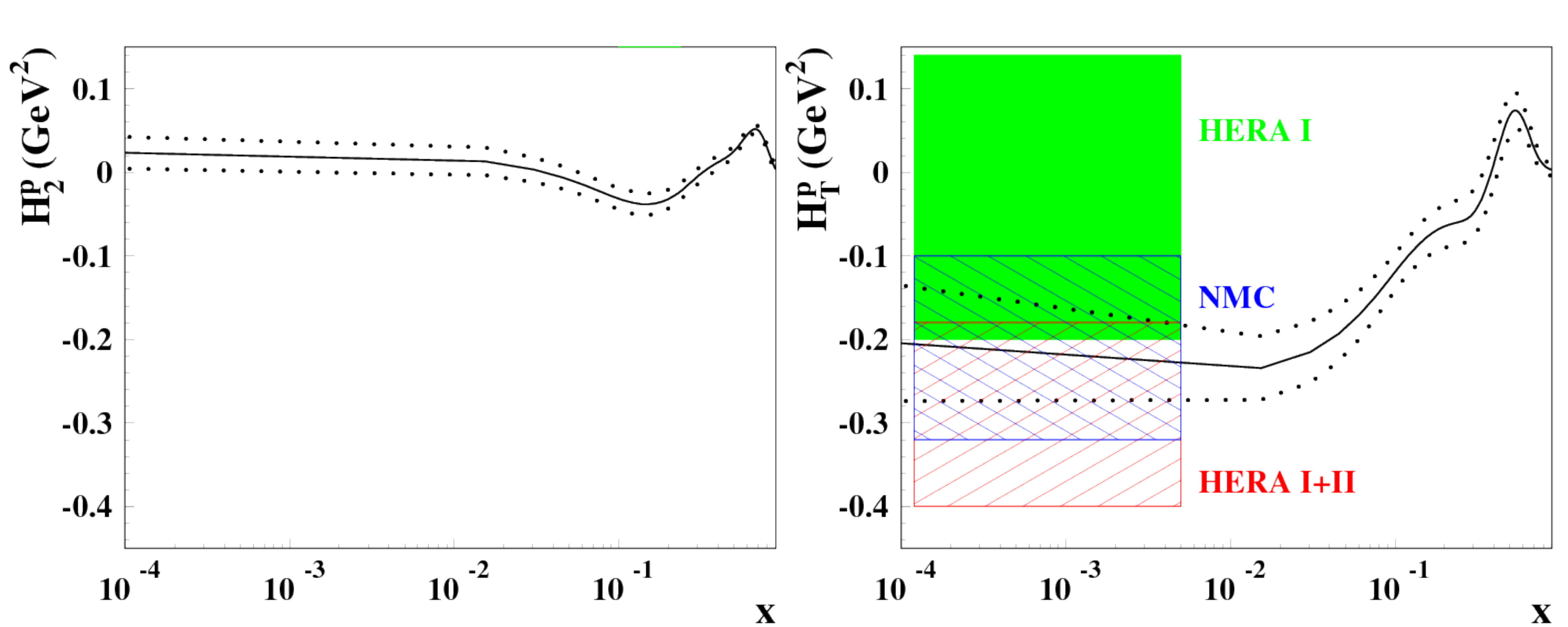}
  \caption{\small
The coefficients of HT contribution to the proton DIS structure function 
$F_2$ (left panel) and $F_T$ (right panel) obtained in the present analysis
(solid line: central value, dots: $1/\sigma$ error band). 
The low-$x$ asymptotic of the HT terms preferred by various 
data sets (shaded area: HERA Run I, left-tilted hatches: NMC, 
shaded area: HERA Run I, right-tilted hatches: HERA Run I+II) 
is given for comparison. 
}
    \label{fig:hts}
\end{figure}

The value of the strong coupling constant $\alpha_s(M_Z)$ is extracted in the present analysis
simultaneously with the PDFs and the heavy-quark masses, similarly to the 
ABM12 case. Basically, it is controlled by the inclusive DIS data 
from the fixed-target experiments and the HERA collider. The values 
preferred by each particular data set are displayed in 
Fig.~\ref{fig:alps} in a chronological order. The most recent data set 
in this row 
stems from a combination of the H1 and ZEUS measurements performed 
during Run~I and Run~II of the HERA collider operation. The 
value of $\alpha_s$ preferred by these data is relatively small as compared to 
the PDF world average~\cite{Agashe:2014kda}, however it is 
somewhat larger than the one obtained from the Run~I HERA combination. As a 
result the NNLO 
value of $\alpha_s(M_Z)=0.1145\pm0.0009$ obtained in the present 
analysis is 
by $1\sigma$ bigger than the earlier ABM12 result based on the 
Run~I HERA data. Earlier DIS data prefer bigger values of $\alpha_s$, 
cf. Fig.~\ref{fig:alps} and they are also more sensitive to the impact 
of higher-twist (HT) contribution to the DIS structure functions. The HT terms 
were studied in detail earlier~\cite{Alekhin:2012ig} and were found to be 
non-negligible for $x\gtrsim 0.01$. In the region of $x\lesssim 0.01$ controlled 
by the HERA and NMC data sets the HT coefficients obtained in the ABM12 analysis
are consistent with zero within uncertainties and therefore were set to zero. 
However, replacement 
of the HERA Run~I data by the Run~I+II ones leads to a change in this 
trend, cf. Fig.~\ref{fig:hts}. As a result the HT coefficients 
in the structure function $F_T$ obtained in the present analysis do not vanish down to $x\sim 10^{-4}$.
At the same time the small-$x$ HT terms in the structure function $F_2$ are 
still compatible with zero. This relation 
implies a pronounced manifestation of the HT 
contribution to the structure function $F_L=F_2-F_T$, in line with the 
earlier observation~\cite{Abt:2016vjh}.

The heavy-quark masses are also considered as free parameters of present fit 
and are determined simultaneously with the PDFs and $\alpha_S$. Throughout we employ 
the running-mass definition which provides an improved perturbative stability as 
compared to the on-shell-mass case~\cite{Alekhin:2013nda,Alekhin:2010sv}.
In this way we obtain the $\overline{MS}$-values of 
$m_c(m_c)=1.252\pm0.018~{\rm GeV}$ and 
$m_b(m_b)=3.83\pm0.12~{\rm GeV}$ driven by the data on 
$c$- and $b$-quark semi-inclusive DIS production, respectively, in combination 
with the inclusive DIS data. Here the heavy-quark contribution 
is described within the framework of fixed-flavor-number (FFN) 
factorization scheme assuming that the heavy quarks are 
produced in the final state. This 
approach provides a good description of existing DIS 
data. Furthermore, the value of $m_c$ obtained is in very good agreement with 
other determinations~\cite{Agashe:2014kda}. This contrasts with the results
obtained in the variable-flavor-number (VFN) approach, which requires 
on-shell masses of $m_c\sim 1.3~{\rm GeV}$ to provide a satisfactory data 
description~\cite{Dulat:2015mca,Harland-Lang:2014zoa,Ball:2014uwa}. 
This value is far below other determinations. Therefore it should be 
rather considered as a tuning parameter somehow curing the VFN scheme effects 
at low $Q^2$ scales~\cite{Accardi:2016ndt}.
The Tevatron and LHC data on the inclusive single-top and top-pair production 
cross section employed in the present analysis allow to determine 
the $t$-quark mass as well. 
A theoretical description of these data is based on the Hathor 
framework~\cite{Aliev:2010zk,Kant:2014oha}
including the NNLO corrections to the $t\bar t$ and the $t$-channel 
single-top production~\cite{Czakon:2013goa,Brucherseifer:2014ama}
and the approximate NNLO corrections 
to the $s$-channel single-top production~\cite{Alekhin:2016jjz}.
The value of $m_t(m_t)=160.9\pm1.1~{\rm GeV}$ is obtained from 
the fit to all existing data including the most recent LHC results for
the collision energy of 13 TeV. The value of $\alpha_s(M_Z)=0.1147\pm0.0008$ 
found with the $t$-quark data added is somewhat larger than the one 
preferred by the DIS data, however, the difference is well within 
the uncertainties. 

In conclusion, we have present an updated version of the ABM12 PDFs with improved 
quark separation and the strange sea determination, tuned to the recent 
data on the DY process, DIS, neutrino-induced charm production, and the 
$t$-quark production collected at the LHC, Tevatron, HERA, and CERN-SPS. 
The values of $\alpha_s$ and the heavy quark masses are determined from 
the fit simultaneously with the PDFs and allow to validate the 
theoretical framework used.

\begin{sidewaystable}[th!]
\renewcommand{\arraystretch}{1.3}
\begin{center}                   
{
\fontsize{8.5}{9.5}\selectfont
\begin{tabular}{|c|c|c|c|c|c|c|c|c|c|c|}   
\hline                           
\multicolumn{2}{|c|}{Experiment}                      
&\multicolumn{2}{c|}{ATLAS}
&\multicolumn{2}{c|}{CMS}  
&\multicolumn{2}{c|}{D0}
&\multicolumn{3}{c|}{LHCb}
\\
\hline                                                    
\multicolumn{2}{|c|}{$\sqrt s$~(TeV)}                      
&7                         
&13
&7  
&8
&\multicolumn{2}{c|}{1.96}
&7
&8
&8                                                        
\\                                                        
\hline
\multicolumn{2}{|c|}{Final states} 
& $W^+\rightarrow l^+\nu$
& $W^+\rightarrow l^+\nu$
& $W^+\rightarrow \mu^+\nu$
& $W^+\rightarrow \mu^+\nu$
&$W^+\rightarrow \mu^+\nu$
& $W^+\rightarrow e^+\nu$
&$W^+\rightarrow \mu^+\nu$
& $Z\rightarrow e^+e^-$                                                        
&$W^+\rightarrow \mu^+\nu$
\\
\multicolumn{2}{|c|}{ }                                          
& $W^-\rightarrow l^-\nu$
& $W^-\rightarrow l^-\nu$
&$W^-\rightarrow \mu^-\nu$
&$W^-\rightarrow \mu^-\nu$
&$W^-\rightarrow \mu^-\nu$
&$W^-\rightarrow e^-\nu$
&$W^-\rightarrow \mu^-\nu$
&                                                         
&$W^-\rightarrow \mu^-\nu$
\\                                                        
\multicolumn{2}{|c|}{ }                                          
& $Z\rightarrow l^+l^-$
& $Z\rightarrow l^+l^-$
&                                                         
&
&
&                                                         
& $Z\rightarrow \mu^+\mu^-$
&                                                         
& $Z\rightarrow \mu^+\mu^-$
\\
\hline                                                    
\multicolumn{2}{|c|}{Cut on the lepton $P_T$ }                      
&$P_T^l>20~{\rm GeV}$      
&$P_T^l>25~{\rm GeV}$      
&$P_T^{\mu}>25~{\rm GeV}$   
&$P_T^{\mu}>25~{\rm GeV}$   
&$P_T^{\mu}>25~{\rm GeV}$
&$P_T^{e}>25~{\rm GeV}$
&$P_T^{\mu}>20~{\rm GeV}$
&$P_T^{e}>20~{\rm GeV}$                       
&$P_T^{e}>20~{\rm GeV}$                       
\\                                                        
\hline                                                    
\multicolumn{2}{|c|}{$NDP$}
&30                      
&6
&11  
&22
&10
&13
&31
&17                            
&32
\\                                                        
\hline
\multirow{2}{4em}{ }
& present analysis
 &31.0
 &9.2
 &{22.4}
 &{16.5}
 &17.6
 &19.0
&45.1
&21.7
 &40.0
\\
\cline{2-11}
 &CJ15
 &--
&--
&--
 &--
 &20
&29
 &--
&--
 &--
\\
\cline{2-11}
 &CT14
 &42
&--
 &--~\footnote{Statistically less significant data with the cut of 
$P_T^{\mu}>35~{\rm GeV}$ are used.}
&-- 
&--
&34.7
 &--
&--
 &--
\\
\cline{2-11}
$\chi^2$ &JR14
 &--
&--
 &--
&--
 &--
&--
 &--
&--
 &--
\\
\cline{2-11}
&{\tt HERAFitter}
 &--
&--
 &--
&--
 &13
&19
 &--
&--
 &--
\\
\cline{2-11}
&MMHT14
 &39
 &--
&--
&--
 &21
&--
 &--
&--
 &--
\\
\cline{2-11}
&NNPDF3.0
 &35.4
&--
 &18.9
 &--
 &--
&--
 &--
&--
 &--
\\
\hline                                          
\end{tabular}
}
\caption{\label{tab:dydata}
\small 
The data on $W$- and $Z$-production in 
$pp$ and $\bar{p}p$ collisions and the 
$\chi^2$ values obtained for these data sets in the present analysis. The 
values obtained in the CJ15~\cite{Accardi:2016qay}, 
CT14~\cite{Dulat:2015mca}, 
JR14~\cite{Jimenez-Delgado:2014twa}, 
HERAfitter~\cite{Camarda:2015zba}, 
MMHT14~\cite{Harland-Lang:2014zoa}, 
and NNPDF3.0~\cite{Ball:2014uwa}
analyses are given for comparison, if available.  
}
\end{center}
\end{sidewaystable}


\begin{thebibliography}{99}

\bibitem{Alekhin:2013nda}
  S.~Alekhin, J.~Bl{\"u}mlein and S.~Moch,
  Phys.\ Rev.\ D {\bf 89} (2014) no.5,  054028
  [arXiv:1310.3059 [hep-ph]].

\bibitem{Alekhin:2015cza}
  S.~Alekhin, J.~Bl{\"u}mlein, S.~Moch and R.~Placakyte,
  arXiv:1508.07923 [hep-ph].

\bibitem{Dulat:2015mca}
  S.~Dulat {\it et al.},
  Phys.\ Rev.\ D {\bf 93} (2016) no.3,  033006
  [arXiv:1506.07443 [hep-ph]].

\bibitem{Harland-Lang:2014zoa}
  L.~A.~Harland-Lang, A.~D.~Martin, P.~Motylinski and R.~S.~Thorne,
  Eur.\ Phys.\ J.\ C {\bf 75} (2015) no.5,  204
  [arXiv:1412.3989 [hep-ph]].

\bibitem{Ball:2014uwa}
  R.~D.~Ball {\it et al.} [NNPDF Collaboration],
  JHEP {\bf 1504} (2015) 040
  [arXiv:1410.8849 [hep-ph]].

\bibitem{Accardi:2016qay}
  A.~Accardi, L.~T.~Brady, W.~Melnitchouk, J.~F.~Owens and N.~Sato,
  Phys.\ Rev.\ D {\bf 93} (2016) no.11,  114017
  [arXiv:1602.03154 [hep-ph]].

\bibitem{Jimenez-Delgado:2014twa}
  P.~Jimenez-Delgado and E.~Reya,
  Phys.\ Rev.\ D {\bf 89} (2014) no.7,  074049
  [arXiv:1403.1852 [hep-ph]].

\bibitem{Camarda:2015zba}
  S.~Camarda {\it et al.} [HERAFitter developers' Team Collaboration],
  Eur.\ Phys.\ J.\ C {\bf 75} (2015) no.9,  458
  [arXiv:1503.05221 [hep-ph]].

\bibitem{Alekhin:2014sya}
  S.~Alekhin, J.~Bl{\"u}mlein, L.~Caminada, K.~Lipka, K.~Lohwasser, S.~Moch, R.~Petti and R.~Placakyte,
  Phys.\ Rev.\ D {\bf 91} (2015) no.9,  094002
  [arXiv:1404.6469 [hep-ph]].

\bibitem{Agashe:2014kda}
  K.~A.~Olive {\it et al.} [Particle Data Group Collaboration],
  Chin.\ Phys.\ C {\bf 38} (2014) 090001.

\bibitem{Alekhin:2012ig}
  S.~Alekhin, J.~Bl{\"u}mlein and S.~Moch,
  Phys.\ Rev.\ D {\bf 86} (2012) 054009
  [arXiv:1202.2281 [hep-ph]].

\bibitem{Abt:2016vjh}
  I.~Abt, A.~M.~Cooper-Sarkar, B.~Foster, V.~Myronenko, K.~Wichmann and M.~Wing,
  Phys.\ Rev.\ D {\bf 94} (2016) no.3,  034032
  [arXiv:1604.02299 [hep-ph]].

\bibitem{Alekhin:2010sv}
  S.~Alekhin and S.~Moch,
  Phys.\ Lett.\ B {\bf 699} (2011) 345
  [arXiv:1011.5790 [hep-ph]].

\bibitem{Accardi:2016ndt}
  A.~Accardi {\it et al.},
  Eur.\ Phys.\ J.\ C {\bf 76} (2016) no.8,  471
  [arXiv:1603.08906 [hep-ph]].

\bibitem{Aliev:2010zk}
  M.~Aliev, H.~Lacker, U.~Langenfeld, S.~Moch, P.~Uwer and M.~Wiedermann,
  Comput.\ Phys.\ Commun.\  {\bf 182} (2011) 1034
  [arXiv:1007.1327 [hep-ph]].

\bibitem{Kant:2014oha}
  P.~Kant, O.~M.~Kind, T.~Kintscher, T.~Lohse, T.~Martini, S.~M{\"o}lbitz, P.~Rieck and P.~Uwer,
  Comput.\ Phys.\ Commun.\  {\bf 191} (2015) 74
  [arXiv:1406.4403 [hep-ph]].

\bibitem{Czakon:2013goa}
  M.~Czakon, P.~Fiedler and A.~Mitov,
  Phys.\ Rev.\ Lett.\  {\bf 110} (2013) 252004
  [arXiv:1303.6254 [hep-ph]].

\bibitem{Brucherseifer:2014ama}
  M.~Brucherseifer, F.~Caola and K.~Melnikov,
  Phys.\ Lett.\ B {\bf 736} (2014) 58
  [arXiv:1404.7116 [hep-ph]].

\bibitem{Alekhin:2016jjz}
  S.~Alekhin, S.~Moch and S.~Thier,
  arXiv:1608.05212 [hep-ph].




 
\end{thebibliography}
\end{document}